\documentclass[12pt]{iopart}

\usepackage{iopams} 
\usepackage{graphics}

\begin{document}

\title[Tunable Functionality and toxicity studies of Titanium Dioxide
Nanotube Layers]{Tunable Functionality and toxicity studies of
  Titanium Dioxide Nanotube Layers}

\author{E. Feschet-Chassot$^1$, V. Raspal$^1$ ,O. K. Awitor$^{1}$,
  F. Bonnemoy$^2$, J.L. Bonnet$^{2,3}$, J. Bohatier$^{2,3}$}

\address{$^1$ Clermont Universit\'e, Universit\'e d'Auvergne, PCSN, BP
  10448, F-63000 Clermont Ferrand}

\address{$^2$ Clermont Universit\'e, Universit\'e Blaise Pascal, UMR
  CNRS 6023, LMGE, BP 10448, F-63000 Clermont Ferrand}

\address{$^3$ Clermont Universit\'e, Universit\'e d'Auvergne,
  Laboratoire de Biologie cellulaire, BP 10448, F-63000
  Clermont Ferrand}

\ead{koawitor@u-clermont1.fr}

\begin{abstract}
  In this work, we have developed economic process to elaborate
  scalable titanium dioxide nanotube layers which show a tunable
  functionality. The titanium dioxide nanotube layers was prepared by
  electrochemical anodization of Ti foil in 0.4~wt$\%$ hydrofluoric
  acid solution. The nanotube layers structure and morphology were
  characterized using x-ray diffraction and scanning electron
  microscopy. The surface topography and wettability was studied
  according to the anodization time. The sample synthesized while the
  current density reached a local minimum displayed higher contact
  angle. Beyond this point, the contact angles decrease with the
  anodization time. Photo-degradation of acid orange 7 in aqueous
  solution was used as a probe to assess the photo-catalytic activity
  of titanium dioxide nanotube layers under UV irradiation. We
  obtained better photocatalitic activity for the sample elaborate at
  higher current density. Finally we use the Ciliated Protozoan
  \textit{T. pyriformis}, an alternative cell model used for in vitro
  toxicity studies, to predict the toxicity of titanium dioxide
  nanotube layers in biological system. We did not observe any
  characteristic effect in the presence of the titanium dioxide
  nanotube layers on two physiological parameters related to this
  organism, non-specific esterases activity and population growth
  rate.
\end{abstract}

\pacs{81.07.-b, 81.16.Rf, 81.07.De}
\vspace{2pc}
\noindent{\it Keywords}: Anodization, Titanium Dioxide Nanotubes,
Contact Angle, Photodegradation, Toxicity. \\
\submitto{Nanotechnology}
\maketitle

\section{Introduction}

The research on developing nanotubes with novel properties by
controlling the nanostructure topography has attracted great interest
because of their variety of applications. In 2001, Gong and co-workers
\cite{DG01} reported the fabrication of vertically oriented highly
ordered $\rm TiO_2$ nanotube arrays up to approximately 500 nm length
by anodization of titanium foil in an aqueous HF electrolyte. Since
then, substantial effort has been devoted to the self organisation and
growth of $\rm TiO_2$ \cite{JM07,OV03,VZ99}. Titanium dioxide nanotube
layers are used as photo-catalysts in water and environmental
purification, as well as, biological and biomedical applications
\cite{KOA08,JM07b,KS06}. In particular, Titanium dioxide nanotubes are
used as a new biomaterial for implants, drug delivery platforms,
tissue engineering and bacteria killing
\cite{KV10,SP10,SCR07,GEA08,IR09,KCP07,CY09}. Another interesting
propriety of $\rm TiO_2$ is its tunable wettability effect
\cite{YS09,EB05}. The ability to modify the surface topography and to
control the wetting behaviour is useful for biomedical
applications. Surface roughness, contact angle, surface energy are the
main factors to understand the biology media and material interaction.
In this work, we present recent results on $\rm TiO_2$ nanotubes
fabricated by anodization of Ti foil in 0.4 wt$\%$ hydrofluoric acid
solution to produce a self-organized porous film structure versus the
anodization time. Such $\rm TiO_2$ nanotube surfaces are of interest
to change the wettability properties of titanium oxide films. The
nanotube layers were characterized using x-ray diffraction and
scanning electron microscopy. We have investigated the surface
wettability of as-anodized samples obtained at different anodization
times and the change in this wettability using octadecylphosphonic
acid (OPDA) coating. The as grown sample synthesized while the current
density reach a local minimum displayed higher contact angle. The
surface of the oxide was covered at this point with a high density of
fine pits. After coating the samples with octadecylphosphonic acid,
the contact angle remained constant. We report on the
photo-degradation of acid orange 7 in aqueous solutions. The acid
orange 7 was used as a probe to assess the photo-catalytic activity of
titanium dioxide nanotube layers under UV irradiation. We obtained
better photocatalytic activity for the sample elaborate at higher
current density. Finally we use the Ciliated Protozoan
\textit{T. pyriformis} to predict the toxicity of titanium dioxide
nanotube layers towards biological system. 

\section{Experimental Details}

\subsection{Sample preparation}

To fabricate anodic $\rm TiO_2$ nanotube layers, we used Ti foil
(Goodfellow 99.6$\%$ purity) with a thickness of 0.1 mm.  The Ti foils
were degreased by successive sonication in trichloroethylene, acetone
and methanol, followed by rinsing with deionized water, dried in the
oven at 100$^{\circ}$C and finally cooled in the
desiccator. Anodization was carried out at room temperature
(20$^{\circ}$C) in 0.4 wt$\%$ HF aqueous solution with the anodizing
voltage maintained at 20 V.

\subsection{Surface characterization}

The surface topography characterization was performed using a Zeiss
Supra 55 VP scanning electron microscope (SEM). The crystalline
structure and phase of the $\rm TiO_2$ nanotube layers were determined
using a Scintag XRD X`TRA diffractometer with Cu $\rm K_\alpha$
radiation.

\subsection{Contact angles}

Surface wettability were investigated with drop shape analysis system
(EasyDrop, Kruss, Hambourg, Germany). The contact angle of 3 $\mu$L
sessile droplet of deionized water was measured on the surface under
ambient conditions. The OPDA solution 50 $\rm \mu$mol/L in toluene was
used to coat the sample. Samples were dipped in the solution for 48 h
and dry in the oven at 70$^{\circ}$C for 24 h.

\subsection{Photo-degradation}

Photo-catalytic experiments were conducted in 3 mL of AO7 solution
(from Acros Organics) with a concentration of $\rm 5.0~10^{-5}$ mol/L,
placed in a cylindrical Pyrex glass reactor. The surface area of the
anodized samples was approximately 3.5 $\rm cm^2$. The glass reactor was
irradiated with polychromatic fluorescent UV lamps (Philips TDL 8 Watt
(total optical power 1.3 Watt), 300 mm long, wavelength range 315-400
nm) in a configuration providing about $\rm 0.9~mW/cm^2$ at the sample
surface. The photo-catalytic decomposition of AO7 was monitored by the
decrease of the solution's absorbance at a wavelength of 485 nm using
a UV-Vis spectrometer (Perkin Elmer Lambda 35).

\subsection{Toxicity assessment}

The potential toxicity of $\rm TiO_2$ nanotube surfaces was evaluated
with \textit{Tetrahymena pyriformis} using two tests previously
validated : inhibition of an enzymatic activity and effect on
population growth rate. For non-specific esterases activities
quantification, a \textit{T. pyriformis} culture in an exponential
growth phase (in PPYS medium) was centrifuged at 300 rpm, and the
supernatant was discarded. The \textit{T. pyriformis} pellet was
suspended in Volvic mineral water. After counting cells under a
microscope, dilution was done to obtain about 4000 cells/mL. 1 mL of
this dilution was incubated for 1h with the different Ti layers at 28
$^{\circ}$C under UV or without UV irradation. After incubation, Ti
layers samples were removed and 1 mL of FDA at 4.8 $\rm \mu M$ was
added (2000 cells/mL in final). Each toxicity test included two
controls : FDA in Volvic water to measure self degradation of this
substrate and FDA with \textit{Tetrahymena pyriformis} (untreated
cells). After 30 min, the fluorescence was measured by a
spectrofluorimeter (Kontron SFM 25, Kontron, Milan, Italy) with a
485~nm excitation filter and a 510 nm emission filter. Experiments
were repeated three times for each sample. To test the inhibition of
development of populations in exponential growth phase, we prepared 8
erlenmeyer flasks (40 mL): 2 for control cultures and 6 for the
samples to test (Ti foil, unannealed $\rm TiO_2$ and $\rm TiO_2$ annealed
at 500 $^{\circ}$C). The samples were deposited at the bottom of the
erlenmeyer flasks and 3 mL were remove at 0h, 3h, 6h and 9h to measure
the optical density (OD at 535 nm).

\section{Results and discussion}

\subsection{$TiO_2$ Nanotube growth process and Layer Characteristics}

The anodization growth was governed by a competition between anodic
oxide formation and chemical dissolution \cite{JM07b} of the oxide as
soluble fluoride complexes according respectively to reactions
(\ref{eq1}) and (\ref{eq2}) :

\begin{eqnarray}
  Ti~+~2~H_2O ~ \rightarrow~ TiO_2~+~4~H^{+}~+~4~e^{-} \label{eq1}\\
  TiO_2~+~4~H^{+}~+~6F^{-}~\rightarrow~[TiF6]^{2-}~+~2~H_2O \label{eq2}
\end{eqnarray}

\begin{figure}[htp]
  \centering
  \includegraphics{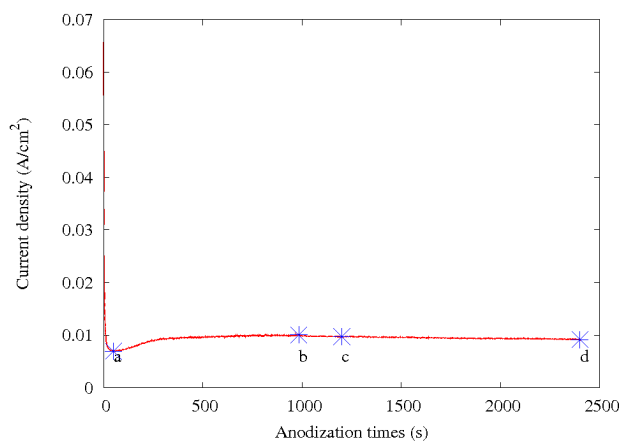}
  \caption{Typical density current time curve for Ti foil
    anodization. Anodization was carried out at room temperature
    (20$^{\circ}$C) in 0.4 wt$\%$ HF aqueous solution with
    the anodizing voltage maintained at 20V.}
  \label{fig:1}
\end{figure}

\begin{figure}[htp]
  \centering 
  \includegraphics{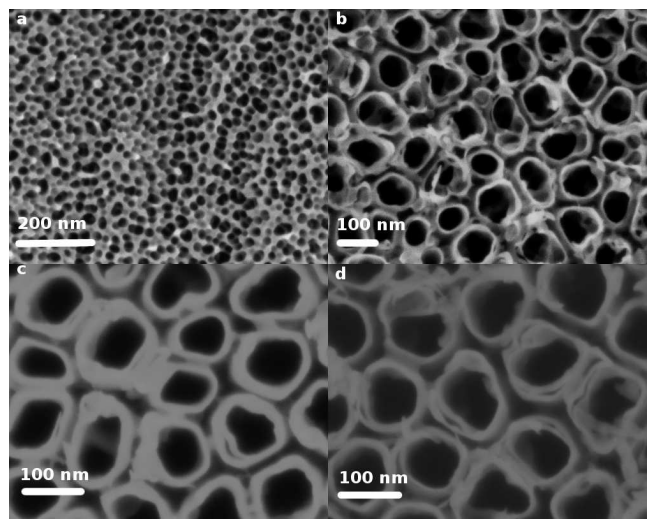}
  \caption{SEM top-view images of samples taken while density current
    was at local minimum (a), local maximum (b) and at the anodization
    time of 20 min and 40 min respectively (c, d).  }
\label{fig:2}
\end{figure}

Figure \ref{fig:1} shows a characteristic density current time curve
for Ti anodization in our operating conditions and figure \ref{fig:2}
shows SEM images of the $\rm TiO_2$ grown at different stages of
growth corresponding to the points a, b, c and d. We can notice that
after an initial exponential decay of the current density to a local
minimum arround 10 $\rm mA.cm^{-2}$ about 70 s. The structure of the
film at this point led to the formation of randomly pits on the oxide
which were shown in figure 2a. The pits were approximately 30 nm in
diameter.

\begin{figure}[htp]
  \centering
  \includegraphics{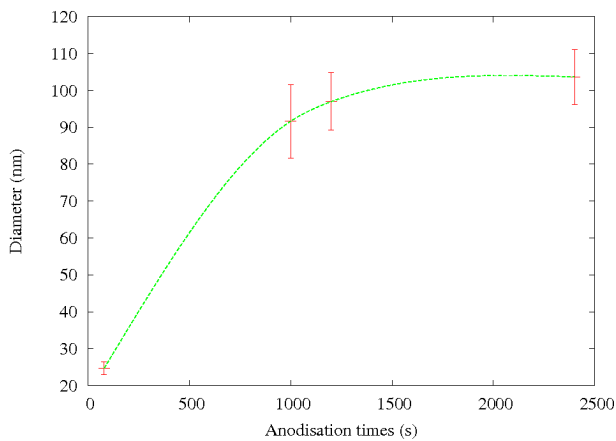}
  \caption{The evolution of the pores diameter of the samples (a),
    (b), (c) and (d).  }
\label{fig:3}
\end{figure}

After the current density increased to a local maximum of 12,7 $\rm
mA.cm^{-2}$ in 1000~s. We have observed ordered nanotube arrays with
approximately 85 nm in diameter as evidenced by the figure
\ref{fig:2}b. After 20 min and 40 min of growth, we observed in figure
\ref{fig:2}c and \ref{fig:2}d ordered nanotube arrays with
approximately 100 nm and 105 nm in diameter respectively. We summarize
in figure \ref{fig:3}, the evolution of the pores diameter. Between 70
and 1000 s, we observe a linear evolution of the diameter versus time
with a high slope. Similarly, the trend is weak between 1000 and 2400
s. At 2400 s, we reach the maximum diameter. This shows that the
dissolution rate of oxide is predominant over the oxide growth
velocity.

\begin{figure}[htp]
  \centering 
  \includegraphics{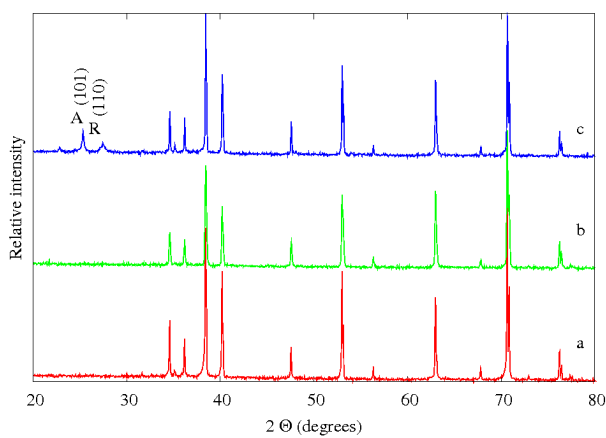}
  \caption{X-ray diffraction patterns of Ti foil (a), as grown during
    40 min (b) and annealed $\rm TiO_2$ nanotube layers at 500$^{\circ}$C
    in oxygen for 2 h (c). Lattice planes indicate anatase (A) and rutile
    (R).  }
\label{fig:4}
\end{figure}

We summarize in figure \ref{fig:4} the x-ray diffraction patterns of
Ti foil and $\rm TiO_2$ nanotube layers anodized for 40 min before and
after annealing at 500$^{\circ}$C in oxygen for 2 h according to the
paper published elsewhere \cite{JM07}.  The unannealed $\rm TiO_2$
nanotube layer exhibits only the peaks from titanium metal foil under
the nanotube layer, while the annealed sample exhibits the main
lattice phases of anatase and rutile (figure~\ref{fig:4}).

\subsection{Contact angles}

\begin{figure}[htp]
  \centering
  \includegraphics{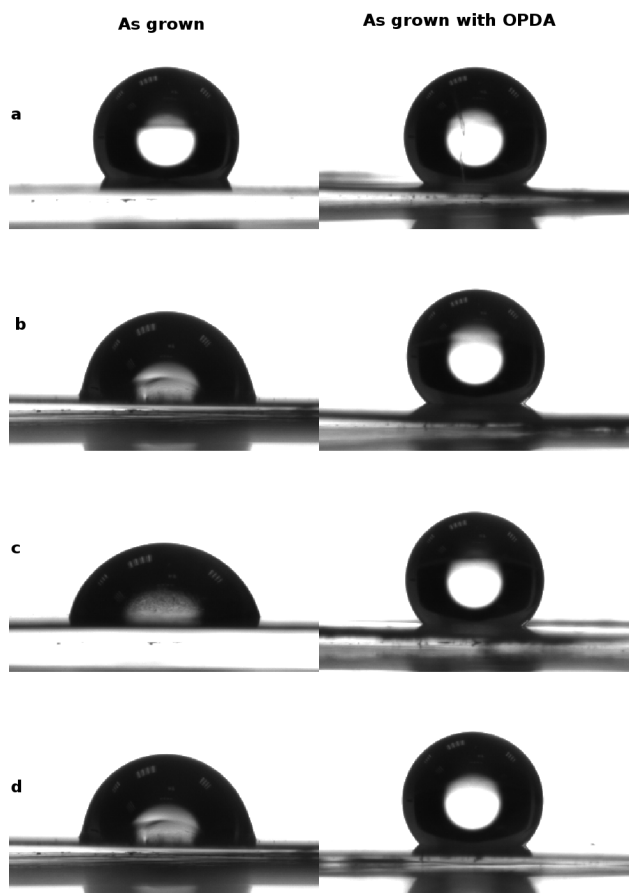}
  \caption{Optical images of water droplets on as grown $\rm TiO_2$
    nanotube layers and after modification with OPDA layer. (a)
    surface at local minimum density current; (b) surface at local
    maximum density current; (c ) surface after 20 min of growth and
    (d) surface after 40 min of growth. }
\label{fig:5}
\end{figure}

Contact angles were measured for each as-anodized sample before and
after octadecylphosphonic acid coating. Each sample was dried in the
oven for 30 min and cooled 15 min in the desiccator before
measurements. We observe in figure \ref{fig:5}, optical images of
water droplets on as grown $\rm TiO_2$ nanotube layers and after
modification with OPDA layer. Results indicated higher contact angle
for the sample covered with pits obtained at local minimum density
current (figure \ref{fig:5}a).

\begin{figure}[htp]
  \centering
  \includegraphics{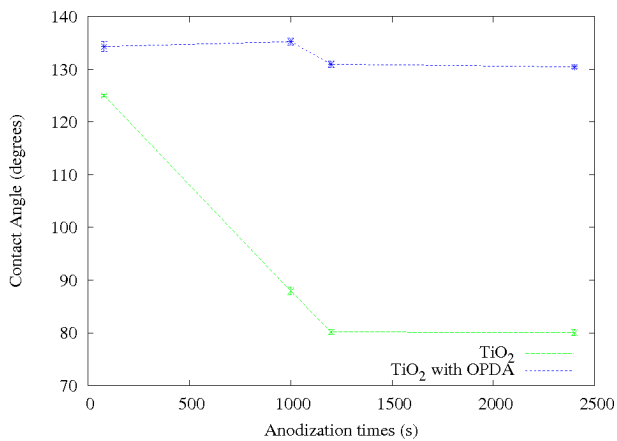}
  \caption{Contact angles vs. Ti foil anodization time.}
  \label{fig:6}
\end{figure}

Beyond this particular point, the contact angles value decrease with
the anodization time. After coating the surface by OPDA, the contact
angle remains the same for all sample. It is superior to 130
$^{\circ}$ which show the hydrophobic behaviour of the OPDA layer
(figure \ref{fig:6}).

\subsection{Photo-catalytic activity measurement}

\begin{figure}[htp]
  \centering
  \includegraphics{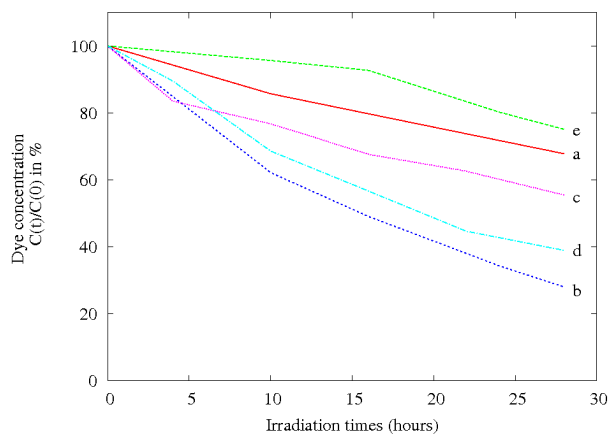}
  \caption{Photo-degradation of acid orange 7 (AO7) dye under UV-lamp
    irradiation at wavelengths of 315-400 nm in the presence of $\rm
    TiO_2$ nanotube layer, as measured by the absorbance of the
    irradiated dye at 485 nm. C(0) is the intitial AO7 ($\rm
    5~10^{-5}~M$) and C(t) is the concentration after time, t, of
    irradiation. (f) unannealed $\rm TiO_2$ nanotube layer; (a), (b),
    (c), (d) $\rm TiO_2$ nanotube layer grown during 70 s, 1000 s, 1200 s,
    2400 s respectively and annealed at 500$^{\circ}$C.}
\label{fig:7}
\end{figure}

The photo-degradation of AO7 in the presence of $\rm TiO_2$ nanotubes
under different conditions is summarized in figure \ref{fig:7}. This
shows the AO7 concentration versus time as determined by the
solution’s absorbance at 485 nm. The initial concentration of the AO7
was $5.0~10^{-5}$ mol/L. C(0) is the initial concentration of AO7
while C(t) is the concentration after time, t, of constant UV
irradiation in 315-450 nm wavelength range. The variation of the
concentration of AO7 in the presence of the $\rm TiO_2$ layer without
irradiation after 5 h is less than 1$\%$. Thus the effect of
adsorption of the dye on the $\rm TiO_2$ surface is negligible. Curve
\ref{fig:7}(e) shows the variation in the concentration of AO7 in the
presence of unannealed $\rm TiO_2$ nanotubes under UV
irradiation. This result indicates that AO7 is not substantially
degraded in the presence of amorphous $\rm TiO_2$ nanotube
layers. Curves \ref{fig:7}(a), \ref{fig:7}(b), \ref{fig:7}(c) and
\ref{fig:7}(d) corresponding to the $\rm TiO_2$ nanotube layers grown
during 70 s, 1000 s, 1200 s, 2400 s respectively and annealed at
500$^{\circ}$C illustrate photo-degradation of AO7. These results show
the decay of organic molecules with UV irradiation in the presence of
the annealed nanotubes. We observed the strongest photo-catalytic
activity for the sample grown during 1200 s.

\begin{figure}[htp]
  \centering
  \includegraphics{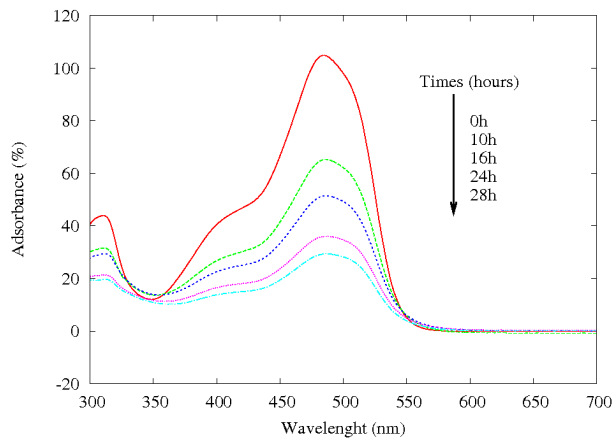}
  \caption{Effect of irradiation with polychromatic light (315-400 nm)
    of AO7 in the presence of $\rm TiO_2$ nanotube layers grown during
    1000 s and annealed at 500$^{\circ}$C on the UV-Vis spectrum in
    the 300-700 nm range.  }
\label{fig:8}
\end{figure}

The effect of irradiation of the sample grown for 1000 s with
polychromatic light (315-400 nm) on the UV-Vis spectrum was showed in
the figure \ref{fig:8}. Absorbance for increasing irradiation time
decreasing from upper curve toward the lower curve.

\subsection{Toxicity tests}

\begin{figure}[htp]
  \centering
  \scalebox{0.45}{\includegraphics{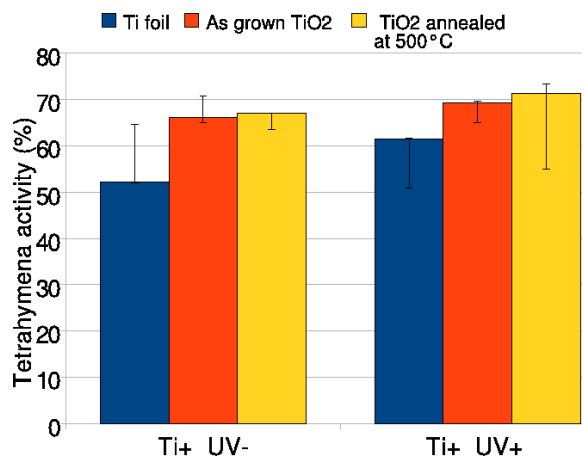}}
  \caption{ \textit{T. pyriformis} activity over the control in the
    presence of Ti foil, amorphous and crystalline $\rm TiO_2$
    nanotube layers without UV irradiation (Ti+UV-) and with constant
    UV light (Ti+UV+) at wavelength range 315-400 nm.}
\label{fig:9}
\end{figure}

Two different tests of toxicity have been made with the Ciliated
protozoan \textit{T. pyriformis} as described in detail elsewhere
\cite{PB01}. This organism is an alternative eukaryotic cell model
including the established fibroblastic cell lines used for in vitro
toxicity studies. All the tests were realized with titanium foil,
amorphous and crystalline $\rm TiO_2$ nanotube layers. The test of
inhibition of non-specific intracellular esterase activity was based
on the hydrolysis of fluorescein diacetate (FDA) by
\textit{T. pyriformis} and quantification of fluorescein released
during 30 min. Esterases are ubiquitous enzymes present in all living
organism and are considered as good biomarkers of well cellular
activities. The aim of this test was to determine the evolution of the
percentage of the \textit{T. pyriformis} activity over the control in
the presence of Ti foils, amorphous and crystalline $\rm TiO_2$
nanotube layers without UV irradiation (Ti+UV-) and with constant UV
irradiation (Ti+UV+) at wavelength 315-400 nm. We can observe an
effect of titanium foil but no significant effect of the amorphous and
crystalline $\rm TiO_2$ nanotube layers. The UV light radiation did
not disturb the results (Figure \ref{fig:9}). Furthermore,
\textit{T. pyriformis} populations growth rate test allows to
integrate different physiological disturbances which could have been
caused by the three types of layers. Growth was followed
photometrically with a measure of optical density ($\lambda$ = 535 nm)
every 3 hours. Reduction in growth compared to a control culture is
indicative of toxicity.

\begin{figure}[htp]
  \centering
  \includegraphics{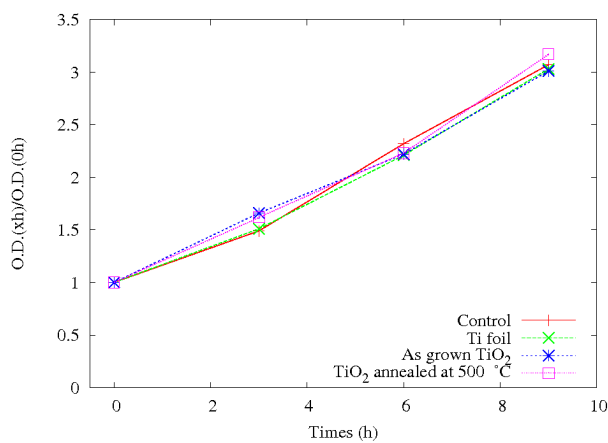}
  \caption{Growing populations of \textit{T. pyriformis} in the
    presence of titanium foil, amorphous and crystalline titanium
    dioxide nanotube layers.}
  \label{fig:10}
\end{figure}

The purpose of this test was to determine a 50$\%$ inhibitory of
growth rate in treated cultures (increase of 50$\%$ of the generation
time compared to a control culture). We did not observe any
characteristic effect related to the inhibition of protozoa's growth
for Ti foils, amorphous and crystalline $\rm TiO_2$ nanotube layers
(Figure \ref{fig:10}).

\section{Conclusions}

We demonstrate the fabrication of controllable as grown surfaces of
titanium dioxide nanotube layers. The contact angles measurements show
clearly the correlation between the surface topography and the surface
wettability. Hydrophobic layer of octadecylphosphonic acid was used to
cap the nanotubes. We show the ability of the titanium dioxide
nanotube layers to degrade the AO7. Such surfaces don't show any
characteristic toxicity effect. In conclusion, we have developed
economic process to elaborate active surfaces of titanium dioxide with
scalable nanotube layers and tunable functionalities.

\section*{References}


\bibliographystyle{iopart-num} 
\bibliography{bibtex_nanotech}


\end{document}